\documentclass[a4paper,11pt]{article}
\usepackage{graphicx}
\usepackage{amsmath}

\title{A gentle introduction to the minimal \\ Naming Game}
\author{Andrea Baronchelli \\ City, University of London }

\date{}

\begin{document}
\maketitle

Social conventions govern countless behaviors all of us engage in every day, from how we greet each other to the languages we speak. But how can shared conventions emerge spontaneously in the absence of a central coordinating authority? The Naming Game model shows that networks of locally interacting individuals can spontaneously self-organize to produce global coordination. Here, we provide a gentle introduction to the main features of the model, from the dynamics observed in homogeneously mixing populations to the role played by more complex social networks, and to how slight modifications of the basic interaction rules give origin to a richer phenomenology in which more conventions can co-exist indefinitely.

\section{Introduction}
The Naming Game~\cite{Steels1995,ng_first} is a multi-agent model in which the individuals perform pairwise interactions to negotiate the conventional forms to be associated with a set of meanings. No central control is available to coordinate the appearance of a globally accepted common lexicon, yet it emerges. How does this happen? Which are the microscopic details allowing for a population-scale agreement?  What is the role of the population size? For example, how does it affect the amount of memory required of the agents or the time needed to reach the final consensus? Moreover, will a consensus even always be reached?

These (and many others) are important questions both from the theoretical point of view and for the applications, but answering them is not easy. Indeed by definition a complex system, such as a community of language users, is an assembly of many interacting (and often simple) units whose collective behavior is not trivially deducible from the knowledge of the rules that govern their mutual interactions. However, predicting the global phenomenology of such a system on the basis of a knowledge of the properties of its elementary constituents is a crucial problem in many fields of research. It is so important, in fact, that there is a specific (sub)discipline entirely devoted to it, namely statistical physics. This is the branch of physics whose goal is to provide the link between macroscopic and microscopic states. For example, it allows us to derive the pressure of an ideal gas (a macroscopic state) starting from the solution to Schr\"{o}dinger's equation for a particle in a box (the microscopic state) \cite{glazer2002statistical,huang1987statistical}. 

Developed to investigate such systems as gases, liquids and solids, statistical physics has proven to be a very fruitful framework also to
describe phenomena outside the classical realm of physics~\cite{loreto07}. In particular, recent years have witnessed the (often successful) attempt to export the concepts and tools developed in the investigation of physical systems for the study of the collective phenomena emerging in social structures~\cite{rmp2009}. Of course, in social phenomena the basic constituents are not particles but humans, and at first sight this could make things much harder. It is a legitimate concern, in fact, but luckily the situation is not as tragic as it might appear. The reason is that humans behave in a surprisingly regular manner. This is true for problems as diverse as mobility patterns \cite{gonzalez2008understanding}, epidemic spreading \cite{pastor2001epidemic}, car traffic dynamics \cite{chandler1958traffic}, and language evolution  \cite{frankfurt}, of course. Thus, depending on the specific issue under consideration, even human beings can be approximated drastically to agents obeying some simple rules. 

A few words are now in order on the issue of modeling. Usually when defining a multi-agent model, the choice is between endowing agents with simple properties, so that one can hope to fully understand what happens in simulations, or with more complicated and realistic structures that yet risk confusing experimental outputs. The statistical physics approach follows the first possibility since it is more interested in
the global behavior of the population. In this perspective its main goal consists in analyzing deeply basic models that can constitute valuable starting points for more sophisticated investigations. Nevertheless, as we shall see, also extremely transparent agents and interaction rules can give rise to very complex and rich global behaviors, and the study of simple models can help to shed light on universal properties. Moreover, it is worth stressing that, quite often, the sociocultural approaches to language evolution lack quantitative investigations, contrary to what happens in the evolutionary approaches \cite{frankfurt}. Later I shall discuss in detail how the main features of the process leading the population to a final convergence state scale with the population size.

This chapter presents and discusses some aspects of the minimal Naming Game~\cite{ng_first} defined by distilling the fundamental ingredients yielding the same global phenomenology observed in robot experiments and more complex models \cite{steels_book,ng_first,Steels1996} and able to reproduce experimental results on the spontaneous emergence of social conventions \cite{centola2015spontaneous}. Due to its simplicity, it has attracted the attention of several researchers in physics, social science, computer science and linguistics. Its formulation is very close in spirit to that of other opinion dynamics models~\cite{fu2008coevolutionary,blythe2009gmc}
(for a detailed analysis of this point see  \cite{rmp2009}). It has been studied in fully connected graphs (i.e. in mean-field or homogeneous mixing populations) \cite{Steels1996,ng_first,Baronchelli_ng_long}, regular lattices \cite{ng_lowdim,lu2008naming}, small world networks \cite{dallasta_ng_smallworld,lu2008naming,liu2009naming}, random geometric graphs \cite{lu2006naming,lu2008naming} and static \cite{dallasta_ng_nets,dallasta_ng_micro,yang2008,baronchelli2011role}, dynamic \cite{nardini2008s,baronchelli2012consensus} and empirical \cite{lu2009naming,trianni2016emergence} complex networks.  It has been shown also that the final state of the system is always consensus~\cite{de2006reach}, but stable polarized states can be reached introducing a simple confidence/trust parameter \cite{baronchelli_ng_trans}. The role of committed minorities in influencing which convention is adopted by the group has also been considered \cite{xie2011social,xie2012evolution,mistry2015committed} 

The Naming Game as defined in~\cite{ng_first,baronchelli_thesis} has also been modified in several ways~\cite{evolang,lu2006naming,baronchelli_ng_trans,wang2007,brigatti2008consequence,lipowski2008bio,lipowski2009,lu2008naming,brigatti2009conventions,lu2009naming,Lei20104046,zhang2010noise,baronchelli2011role}, representing the fundamental core of more complex models in computational cognitive sciences \cite{cg_pnas,baronchelli10}. Furthermore, from the point of view of the applications, its relevance in system-design has been pointed out in the context of sensor networks \cite{akyildiz2002survey}, in relation to such problems as autonomous key creation or selection for encrypted communication  \cite{lu2008naming} and, more recently, as a tool for investigating the community structure of social networks \cite{lu2009naming,zhang2010noise}.

The constraints of this chapter neither permit nor necessitate a detailed review of all of the above mentioned results. Rather, it focuses upon some key aspects of the model dynamics in order to illustrate which kind of benefits can be obtained from a statistical physics approach to the modeling of language games. We will start defining the model and discussing its basic phenomenology. We will then inspect the role played by the population structure, looking at different cases in which the set of possible interactions of each individual is limited to a fixed number of neighbors. We shall see that the statistics of this underlying interaction patterns dramatically affects the global dynamics, both from the point of view of the time needed to reach consensus and from the memory required to the agents. We will then look at two slightly modified versions of the minimal NG, that will deepen our understanding of the convergence process and the role played by inter-agent feedback, respectively.

\section{The Model}
The minimal Naming Game (NG) \cite{ng_first,Baronchelli_ng_long} is played by a population of $N$ agents that engage in pairwise interactions in order to {\em negotiate} conventions (i.e., associations between forms and meanings), and it is able to describe the emergence of a global consensus among them.  For the sake of simplicity the model does not take into account the possibility of homonymy, so that all meanings are independent and one can work with only one of them, without loss of generality. An example of such a game is that of a population that has to reach the consensus on the name (i.e., the form) to assign to an object (i.e., the meaning) exploiting only local interactions, and we will adopt this perspective in the remainder of this paper. 

\begin{figure}[t]
\centerline{
\includegraphics[width=0.8\columnwidth]{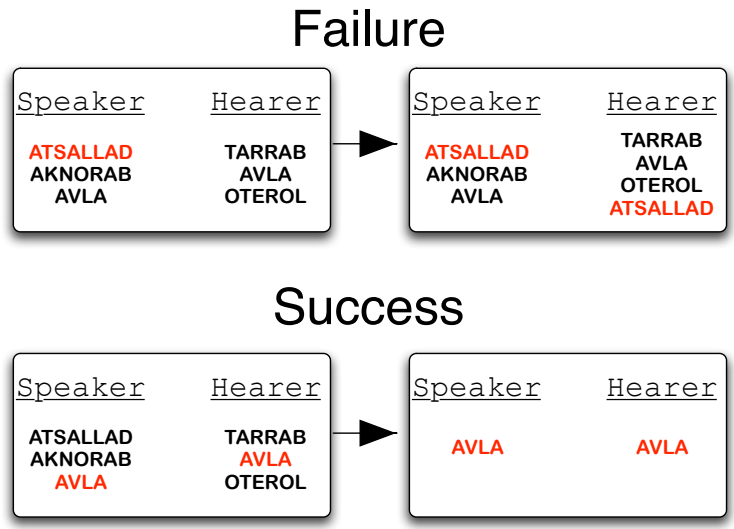}}
  \caption{{\footnotesize Naming Game. Examples of the dynamics of the
    inventories in a failed (top) and a successful (bottom) game. The
    speaker selects the highlighted word by random extraction. If the hearer does
    not possess that word he includes it in his inventory
    (top). Otherwise both agents erase their inventories, keeping only
    the winning word (bottom). 
    \label{rules_naming}}}
    \vspace{0.5cm}
\end{figure}

Each agent disposes of an internal inventory, in which an a priori
unlimited number of words can be stored. As initial conditions we
require all inventories to be empty. At each time step ($t=1,2,..$), a
pair of neighboring agents is chosen randomly, one playing as
``speaker'', the other as ``hearer'', and negotiate according to the
following rules (see Fig.~\ref{rules_naming}):
\begin{itemize}
\item the speaker randomly selects one of its words (or invents a new word if its inventory is empty) and conveys it to the hearer; 
\item if the hearer's inventory contains such a word, the two agents
  update their inventories so as to keep only the word involved in the
  interaction ({\em success});
\item otherwise, the hearer adds the word to those already stored in its inventory ({\em failure}).
\end{itemize}

\noindent With this scheme of interaction, the assumption of the absence of
ho\-mon\-y\-my simply translates into assuring that each newly
invented word had never appeared before in the population. Thus,
single objects are independent (i.e., it is impossible that two agents
use the same word for two different objects), and their number becomes
a trivial parameter of the model. 
For this
reason, as we mentioned above, we concentrate on the presence of one
single object, without loss of generality.

It is also interesting to note that the problem of homonymy has been
studied in great detail in the context of evolutionary game theory, and
it has been shown~\cite{KomarovaNiyogi2004} that languages with
homonymy are not evolutionary stable. However, it is obvious that
homonymy is an essential aspect of human languages, while synonymy
seems less relevant. The authors solve this apparent paradox by
noting that if we think of ``words in a context''~homonymy almost
disappears while synonymy acquires a much greater role. In the
framework of the minimal NG, homonymy is not always an unstable feature (see
\cite{cg_pnas} for an example), and its
survival depends in general on the size of the meaning and signal
spaces~\cite{gosti07}. This observation also fits well with the implicitly assumed
inferential model of learning, according to which we assume that
agents are placed in a common environment and they are able to point to
referents. Subsequently, after a failure, the speaker is able to point out the named
object (or referent) to the hearer which in its turn can assign the
new name to it.

Another important remark concerns the random extraction of the word in
the speaker's inventory. Many previously proposed models attempted to 
give a more detailed representation of the
negotiation interaction assigning weights to the words in the
inventories. In such models, the word with the largest weight is
automatically chosen by the speaker and communicated to the
hearer. Success and failures are translated into updates of the
weights: the weight of a word involved in a successful interaction is
increased to the detriment of those of the others (with no deletion of
words); a failure leads to the decrease of the weight of the word not
understood by the hearer.  An example of a model including weights
dynamics can be found in~\cite{lenaerts2005} (and references therein).
For the sake of simplicity the minimal NG described above
avoids the use of weights. Indeed, weights are apparently more
realistic form a cognitive point of view, but their presence is not essential for the emergence of a
global collective behavior of the system~\cite{baronchelli_thesis}.

\section{Basic Phenomenology}
The non-equilibrium dynamics of the minimal NG is characterized by three
temporal regions: (1) initially the words are invented; (2) then they
spread throughout the system inducing a reorganization process of the
inventories; (3) this process eventually triggers the final
convergence towards the global consensus (all agents possess the same
unique word).

More precisely, the main quantities that describe the system's evolution are~\cite{ng_first}:  
\begin{itemize}
\item the total number $N_{w}(t)$ of words in the system at the time
$t$ (i.e., the total size of the memory);
\item the number of different words $N_{d}(t)$ in the system at the time $t$;
\item the average success rate $S(t)$,
i.e., the probability, computed averaging over many simulation runs,
that the chosen agent gets involved in a successful interaction at a
given time $t$.
\end{itemize}

\begin{figure}[t]
\begin{center}
\includegraphics[width=0.8\columnwidth]{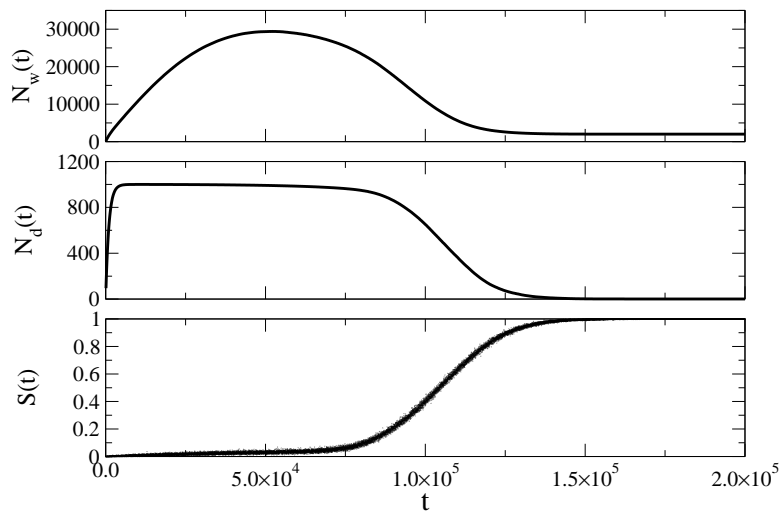}
\end{center}
\caption{{\footnotesize Time evolution of the most relevant global
properties of the Naming Game. From up to down: the total number of
words, $N_w(t)$, the number of different words known by the agents,
$N_d(t)$, and the probability of a successful interaction at a give
time, $S(t)$. Convergence is reached with a quite abrupt
disorder/order transition that starts approximately just after the
peak of the $N_w(t)$ curve has disappeared. Data are relative to a
population of $N=2000$ agents and averaged over $300$ simulation
runs. \label{f:classic2k}}}
\vspace{0.5cm}
\end{figure}

The consensus state is obtained when $N_{d}=1$ and $N_{w}=N$ (so that
$S(t)=1$), and the temporal evolution of the three main quantities is
presented in Fig.~\ref{f:classic2k}.  First, many
disjoint pairs of agents interact, with empty initial inventories:
they invent a large number of different words ($N/2$, on average) that
start spreading throughout the system through failure events. 
Indeed, the number of words decreases only by means of successful
interactions, which are limited in the early stages by a very low
overlap between inventories.  The number of different words
$N_{d}$ grows, rapidly reaching a maximum, and then saturates to a plateau 
where $N_{d}=N/2$, on average. The total number $N_{w}$ of words, on the other hand,
keeps growing after $N_{d}$ has saturated, since the words continue to
propagate throughout the system even if no new one is introduced.  In the
subsequent dynamics, strong correlations between words and agents
develop, driving the system to a final rather fast convergence to the
absorbing state. The
S-shaped curve of the success rate in Fig.~\ref{f:classic2k} summarizes the
dynamics: initially, agents hardly understand each others ($S(t)$ is
very low); then the inventories start to present significant overlaps,
so that $S(t)$ increases until it reaches $1$. It is worth noting that
the established communication system is not only effective (agents can
understand each others) but also efficient (no memory is wasted in the
final state).

\begin{figure}[t]
\centerline{
 \includegraphics[width=0.8\columnwidth]{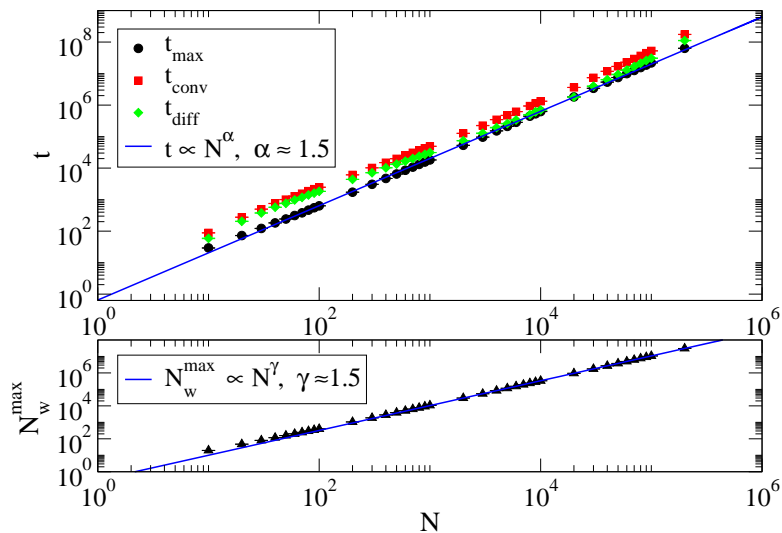}}
 \caption{{\footnotesize Naming Game. (Top) scaling of the peak and convergence time,
   $t_{max}$ and $t_{conv}$ along with their difference, $t_{diff}$.
   All curves scale with the power law $N^{1.5 \pm 0.1}$. (Bottom) the maximum
   number of words obeys the same power law
   scaling.}\label{scaling_N_naming}} 
   \vspace{0.5cm}
\end{figure}

\subsection{Scaling Relations}

In the dynamical evolution described above we can point out at least two crucial points. The first is the time $t_{max}$ at which the total number of words in the system $N_w(t)$ reaches its peak. The height of this peak, $N_w^{max}$, corresponds indeed to the maximum amount of memory required by each agent, $N_w^{max} / N$, during the whole process and it is therefore important.
The second relevant instant is the convergence time $t_{conv}$, at which the dynamics ends for all practical purposes. To these we can add the time span that separates these two moments, i.e. $t_{diff} = t_{conv} - t_{max}$. 

It would have been very easy to determine these quantities from the simulations that produced the curves showed in Fig.~\ref{f:classic2k}, concerning a population made of $N=2000$ individuals. Yet they would have been almost meaningless numbers telling us how many simulation steps are needed on average for this particular population size. It is therefore much more interesting to focus on how the relevant quantities \textit{scale with the system size}, i.e. to look at how they are related to the number of individuals. Interesting questions are therefore of the form: what happens if the population size is, say, doubled? Is the convergence time, for example, doubled too? Or rather does it become much slower? This is a typical way of addressing problems in statistical physics (see, for instance, \cite{sethna2006statistical}). It has profitably been exported to study the minimal NG, for which it has been found that (see Fig. \ref{scaling_N_naming}):

\begin{equation}
 t_{conv} \sim t_{max} \sim t_{diff} \sim N^{\alpha} \;\;\; \mbox{with } \;\;\; \alpha \approx 1.5,
 \label{e:times}
\end{equation}
and 
\begin{equation}
 N_w^{max} \sim  N^{\gamma} \;\;\; \mbox{with } \;\;\; \gamma \approx 1.5.
\end{equation}

\noindent These relations can be recovered also through analytical scaling arguments \cite{ng_first,Baronchelli_ng_long}, and their implications are profound. In particular, the scaling of $N_w^{max}$ implies that the average amount of memory required to each agent is $N_w^{max} / N \sim N^{1/2}$. Thus, the cognitive effort an agent has to take, in terms of maximum inventory size,  \textit{depends on the system size} and, in particular, diverges as the population gets larger and ideally goes to infinity (the so-called thermodynamic limit in the language of statistical physics). Concerning the time to reach convergence, $t_{conv}$, as well as the other convergence times, for now we can only acknowledge the results reported in Eq.~(\ref{e:times}). 

A natural question is now what the scaling relations we have found depend on. Likewise, we can ask whether they can be modified. The answer is not univocal. On one level, of course, the behavior just described depends on many of the details of the introduced model. It is possible in principle, and it is actually the case \cite{baronchelli2011role} some of them are irrelevant in this context, but in general the adopted modeling scheme matters. On another, deeper level, however, we can ask whether there are any features that, without changing the microscopic interaction rules, yield a different population-scale phenomenology. We shall see in the next paragraph that at least one such feature exists, and it is the interaction pattern underlying the pairwise communications of the individuals.

\section{The role of topology}

In the previous section we introduced the minimal NG model prescribing that, at each time step $t=1,2,..$, two agents are \textit{randomly} selected. The assumption behind this homogeneous mixing, or ``mean-field", rule is that the population is not structured and that any agent can in principle interact with any other. In general, however, this is not true, and the \textit{topology} on which the population is embedded identifies the set of possible interactions among the individuals.  Thus, the group of communicating individuals can be described as a \textit{network} in which each node represents an agent and the links connecting different nodes determine the allowed communication channels. The (statistical) properties of the underlying network can therefore affect the overall dynamics of the model. In the coming section we will see that this is actually the case.  

Recent years have witnessed the birth and fast development of the field of complex networks~\cite{barabasi02,satorras_libro,caldarelli_book,barratbook}. First of all, it was realized that a schematization in terms of nodes and links representing their interactions was a powerful tool to describe and analyze a large set of different systems, belonging to technological (Internet, the web, etc.), natural (food webs, protein interaction networks, etc.) or social (networks of scientific collaborations, acquaintances, etc.) domains. Surprisingly, it was then found that almost all the investigated systems share a certain number of peculiar and completely unexpected properties, which were not captured by the models known up to that moment. For example, human social networks are highly heterogeneous, where most people have a relatively small number of acquaintances and where only a few social hubs are hugely connected \cite{caldarelli_book}. From our point of view, such complex networks and the artificial attempts to reproduce them constitute possibly the most natural interaction patterns to study the NG, but we will see that it is convenient to start studying the effect of simpler topologies first and then move to the most complex ones.

We discuss below the main findings obtained in embedding the minimal NG on different structured topologies of increasing complexity. 
The analysis in unavoidably somewhat technical, but the reader who is not familiar with complex network theory need not worry. Indeed, it turns out that no matter how complex the underlying topology is, the properties that affect the global behavior of the system are essentially two, namely the finite connectivity and the small-world property\footnote{We do not consider here the effect of such features as strong clustering or community structures,  concerning which we refer the interested reader to \cite{dallasta_ng_nets}.}. The first refers to the fact that a given agent can interact only with a fixed subset of the whole population. The latter describes the evidence that the average distance length $\langle l \rangle$ between pair of nodes is ``very small''. More precisely, $\langle l \rangle$ scales logarithmically, or slower, with the system size. This property is of course absent in regular structures, where $\langle l \rangle \sim N^{1/d}$, $d$ being the dimensionality of the system. Their impact can be summarized as follows:

\begin{table}[t]
\begin{center}
\begin{tabular}{l|ccc}

  & \hspace{0.2cm} Mean-field  \hspace{0.2cm} & \hspace{0.2cm} Lattices ($d \leq 4$)  \hspace{0.2cm}& \hspace{0.2cm} Networks  \hspace{0.2cm}\\ 
  \hline
  & & & \\
  Maximum memory  & $N^{1.5}$ & $N$ & $N$ \\ 
  & & & \\
  Convergence time  & $N^{1.5}$ & $N^{1+\frac{2}{d}}$ & $N^{1.4 \pm 0.1}$ \\
  & & & \\
\end{tabular}
\caption{{\footnotesize Scaling with the system size $N$ of the maximum number of
  words (memory) and time of convergence. Networks, thanks to the
  small-world property and the finite connectivity, ensure a trade-off
  between the fast convergence of mean-field topology and the small
  memory requirements of lattices. }}
\label{t:summary}
\end{center}
\vspace{0.5cm}
\end{table}

\begin{enumerate}
\item Finite connectivity implies finite memory requirements to the agents, disentangling the maximum inventory size from the number of individuals in the population.
\item The small-world property guarantees ``fast" convergence, allowing the fast spreading of words created in otherwise far-apart regions of the underlying topology.
\end{enumerate}

\noindent In Point 2, ``fast" convergence means the fastest scaling of the convergence time observed in all the numerical experiments conducted so far. Table \ref{t:summary} recapitulates the results for a different kind of topologies. The mean-field population is of course small-world (the distance between any pair of agents is simply $1$), but as we have seen before the non-finite connectivity implies a diverging memory requirement. Low-dimensional regular lattices assure finite connectivity but lack the small-world property, hence the memory per agent is finite (since the global memory requirement scales just as the population size $N$), but convergence is extremely slow ($t_{conv} \sim N^3$ for $d=1$). Finally, complex networks exhibit both properties and are therefore the most advantageous arrangement (as well as the more natural one), assuring at the same time finite memory and ``fast" convergence. We analyze below the different scenarios in more detail.

\subsubsection*{Low dimensional lattices}

Before looking at the effects of an underlying complex topology it seems reasonable, as mentioned above, to  look at the the effects of simpler situations, namely low dimensional regular lattices \cite{baronchelli_ng_lowdim} . Moreover,  $d-$dimensional lattices have traditionally been used as underlying topologies of many classical models of statistical physics, and there are well established methods to tackle them~\cite{marro}. Here, the number of neighbors is finite, the structure is regular, and there is a complete homogeneity among the agents. In the minimal NG, low dimension grids induce a coarsening dynamics, so that the time required by the system to converge is much slower. On the other hand, finite connectivity keeps the memory required to each agent finite.

\begin{figure}[t]
\centerline{
\includegraphics*[width=0.8\columnwidth]{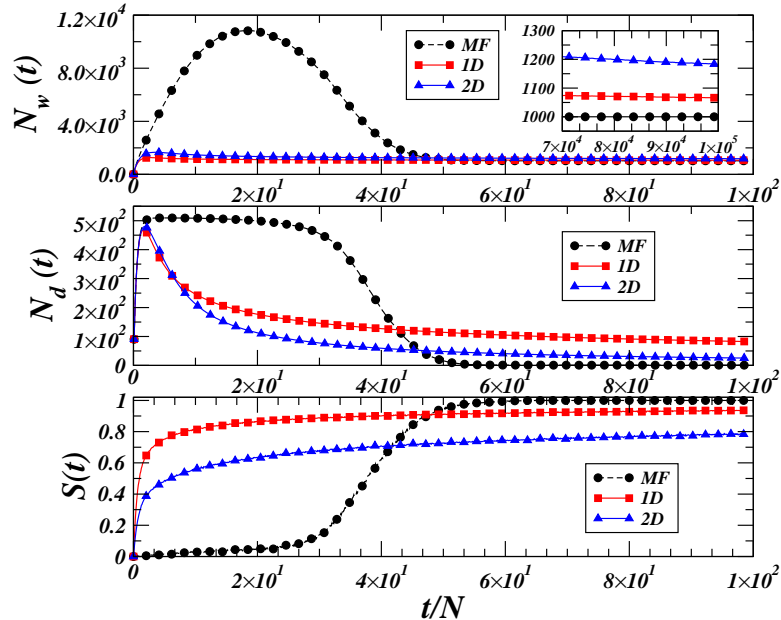}}
\caption{{\footnotesize Evolution of the total number of words $N_{w}$ (top), of the
  number of different words $N_{d}$ (middle), and of the average
  success rate $S(t)$ (bottom), for a fully connected graph
  (mean-field, MF) (black circles) and low dimensional lattices ($1D$,
  red squares and $2D$, blue triangles) with $N=1024$ agents, averaged
  over $10^3$ realizations. The inset in the top graph shows the very
  slow convergence for low-dimensional systems.}}
  \vspace{0.5cm}
\label{pheno_naming_lowdim}
\end{figure}

On low-dimensional lattices each agent can rapidly interact two or more times
with its neighbors, favoring the establishment of a local consensus
with a high success rate (Fig.~\ref{pheno_naming_lowdim}, red squares
for $1D$ and blue triangles for $2D$), namely of small sets of
neighboring agents sharing a common unique word. Later on these
``clusters'' of neighboring agents with a common unique word undergo a
coarsening phenomenon~\cite{baronchelli_ng_lowdim} with a competition among
them driven by the fluctuations of the interfaces. The
coarsening picture can be extended to higher dimensions, and the
scaling of the convergence time has been conjectured as being ${\cal
  O}(N^{1+1/d})$, where $d\le4$ is the dimensionality of the space.
This prediction has been checked numerically. On the other hand the
maximum total number of words in the system (maximal memory capacity)
scales linearly with the system size, i.e., each agent needs only a
finite memory. 

\subsubsection*{Small-world networks}

Results concerning the mean field case, on the one hand, and regular structures, on the other, act as fundamental references to understand the role of the different properties of complex networks. We start by addressing the role of the small-world property (short average distance between any pair of nodes), which is one of the most characteristic features of many different networks. In particular we focus on a model, proposed by Watts and Strogatz~\cite{watts_strogatz,watts_book}, which allows to pass progressively from regular structures to random graphs by tuning the $p$ parameter describing the probability that a link of the regular structure is rewired to a random destination. The main result is that the presence of shortcuts, linking agents otherwise far from each other, allows to recover the fast convergence typical of the mean-field case \cite{dallasta_ng_smallworld}. The finite connectivity, on the other hand, keeps the demanded amount of memory finite, as in regular structures. 

Studying the dynamics of the minimal NG in small-world lattices, two different regimes are observed. For times shorter than
a cross-over time, $t_{cross} = {\cal O}(N/p^2)$, one observes the
usual coarsening phenomena as long as the clusters are typically
one-dimensional, i.e., as long as the typical cluster size is smaller
than $1/p$. For times much larger than $t_{cross}$, the dynamics is
dominated by the existence of short-cuts and enters a mean-field like
behavior. The convergence time scales therefore as $N^{3/2}$
and not as $N^{1+1/d}$ (as in low-dimensional lattices). As anticipated above, small-world topology allows thus to
combine advantages from both finite-dimensional lattices and
mean-field networks: on the one hand, only a finite memory per node is
needed, in opposition to the ${\cal O}(N^{1/2})$ in mean-field; on the
other hand the convergence time turns out to be much shorter than in
finite dimensions. 

\begin{figure}[t]
\centerline{
\includegraphics*[width=0.8\columnwidth]{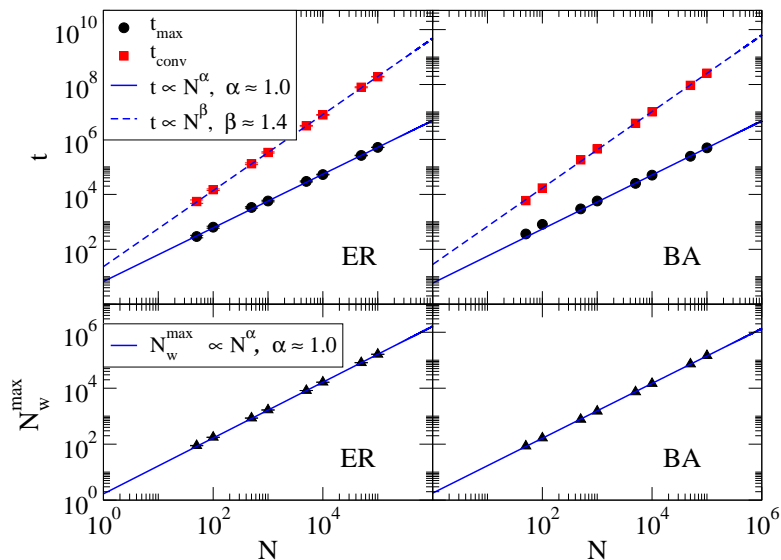}}
\caption{{\footnotesize Top: scaling behavior with the system size $N$ for the time
  of the memory peak ($t_{max}$) and the convergence time ($t_{conv}$)
  for ER random graphs (left) and BA scale-free networks (right) with
  average degree $\langle k\rangle=4$. In both cases, the maximal
  memory is needed after a time proportional to the system size, while
  the time needed for convergence grows as $N^{\beta}$ with $\beta
  \simeq 1.4 \pm 0.1$. Bottom: In both networks the necessary memory capacity
  (i.e. the maximal value $N_w^{max}$ reached by $N_w$) scales
  linearly with the system size.\label{scaling_naming_networks}}}
  \vspace{0.5cm}
\end{figure}

\subsubsection*{Complex Networks}

In~\cite{dallasta_ng_nets} most of the relevant features exhibited by complex networks have been explored systematically, mainly by means of computer simulations. It must also be noted that the (minimal) NG  as described above is not well-defined on general networks. When the degree distribution is heterogeneous, it does matter if the first randomly chosen agent is selected as a speaker and one of its the neighbor as the 
hearer or vice versa: high-degree nodes are in fact more easily chosen as neighbors than low-degree vertices. Several variants of the Naming Game on generic networks can be defined. In the {\em direct Naming Game} ({\em reverse Naming Game}) a randomly 
chosen speaker (hearer) selects, again randomly, a hearer (speaker) among its neighbors. In a {\em neutral} strategy one selects an edge and assigns the role of speaker and hearer with equal probability to one of the two nodes~\cite{dallasta_ng_nets}.

Here we only report on the global behaviour of the system (direct NG), and we refer
to~\cite{dallasta_ng_nets} for an extensive discussion.
Fig.~\ref{scaling_naming_networks} shows that the convergence time
$t_{conv}$ scales as $N^{\beta}$ with $\beta \simeq 1.4 \pm 0.1$, for
both Erd\"{o}s-Renyi (ER)~\cite{erdos59,erdos60} random graphs (where the degree distribution is peaked and
all the nodes have very similar connectivity patterns) and Barabasi-Albert
(BA)~\cite{barabasi99} networks (that have a power
law degree distribution given by $P(k) \sim k^{- 3}$, so that the
vast majority of the nodes is poorly connected while few hubs have large
degrees). The scaling laws observed for the
convergence time is a general robust feature that is not affected by
further topological details, such as the average degree, the
clustering or the particular form of the degree distribution. The
value of the exponent $\beta$ has been checked for various $\langle
k\rangle$, clustering, and exponents $\gamma$ of the degree
distribution $P(k)\sim k^{-\gamma}$ for scale-free networks
constructed with the uncorrelated configuration model
(UCM)~\cite{molloy_reed95,catanzaro05}.

\subsection{Microscopic dynamics}
\begin{figure}[ht]
\centerline{ 
\includegraphics*[width=\textwidth]{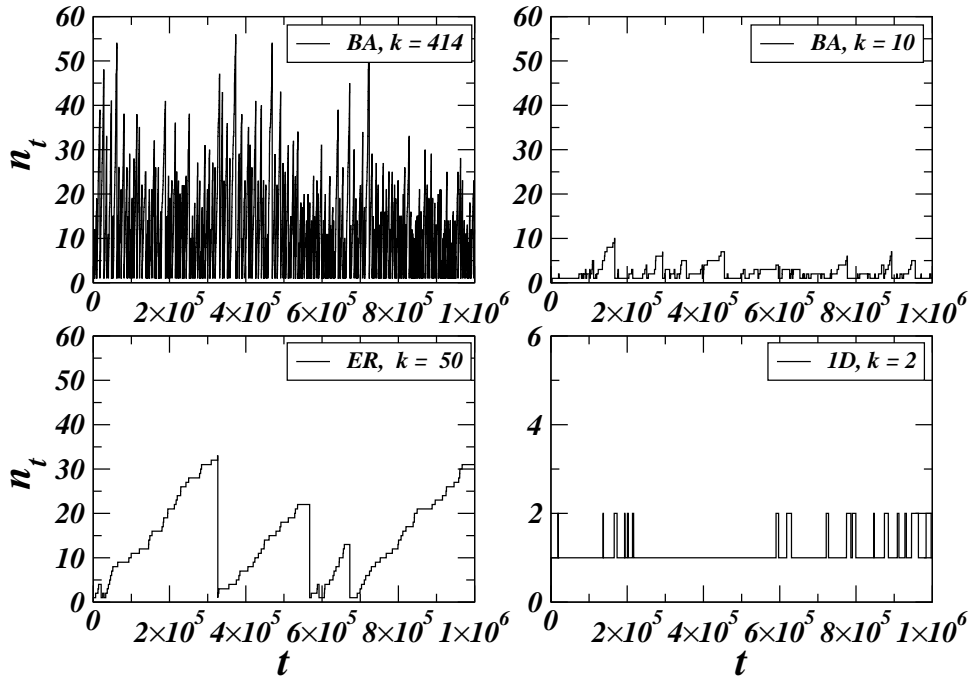} }
\caption{{\footnotesize Temporal series of the inventory size of a single agent
  in different topologies. Top: Series from a Barab\'asi-Albert (BA)
  network with $N=10^4$ nodes and average degree $\langle k\rangle
  =10$, for nodes of high degree (e.g. $k=414$) and low degree
  (e.g. $k=10$). Bottom: Series for nodes in Erd\"os-R\'enyi random
  graph ($N=10^4$, $\langle k \rangle = 50$) and in a one-dimensional
  ring ($k=2$).}}
  \vspace{0.5cm}
\label{fig_micro}
\end{figure}

Along with the global quantities we have studied so far, it is also
interesting to investigate the microscopic activity patterns of single
agents, and to study how they are affected by the underlying
topology~\cite{dallasta_ng_micro}. In particular, in complex networks,
simple properties of the degree distribution (namely the first two
moments) turn out to dramatically affect the memory requirements of
the agents, in a way that depends both on the general features of the
considered network and on the connectivity of the single
agents. Without entering in the mathematical details that allow 
for precisely quantifying the impact of topology on agents
activity~\cite{dallasta_ng_micro}, a simple look at
Figure~\ref{fig_micro} permits a qualitative idea of the
relevance of the phenomenon to be gained. Here the time evolution of the inventory
size of single agents is presented, and the role of connectivity
patterns is evident. Top panels refers to a highly connected node (i.e an ``hub") (left) and a less
connected node (right) belonging to the same Barab\'asi-Albert
network~\cite{barabasi_albert_first}. The bottom left panel, on the other hand, concerns the
activity of an average node on a homogeneous Erd\"os-R\'enyi random
graph~\cite{erdos1959}. Finally, the bottom right
square presents the activity of an agent belonging to a
population arranged on the nodes of a linear chain, whose inventory
never exceeds the size of two words. In summary, the microscopic point
of view not only supports and complements the study of global
quantities, but also allows deeper connections between
the learning process of the agents (i.e., the dynamics of acquisition
and deletion of words of a single agent) and the topological
properties of the system to be pointed out.

\begin{figure}[t]
\centerline{
\includegraphics*[width=\columnwidth]{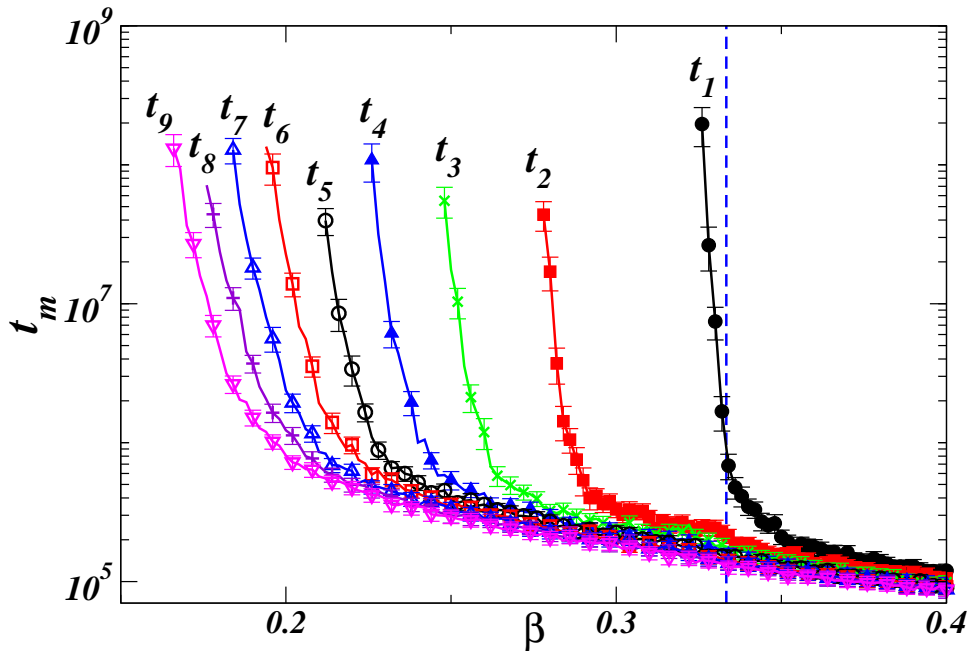}}
\caption{{\footnotesize Time $t_{x}$ required to a population on a fully-connected
  graph to reach a (fragmented) active stationary state with $x$
  different opinions. For every stable number of surving words $m > 2$, the time $t_{m}$ diverges at
  some critical value $\beta_{c}(m) < \beta_{c}$. }}
  \vspace{0.5cm}
\label{fig_transitions}
\end{figure}

\section{Lessons from slightly modified NGs}

As mentioned above, each detail of the microscopic rules of a multi-agent model has a potentially dramatic effect on the observed macroscopic, population-scale dynamics. In this section, we survey two examples in which slight changes in the minimal NG rules yield interesting results. In both cases the rule under inspection is the one defining the update scheme that the agents have to follow after a successful interaction. In the first setting, the introduction of a simple parameter allows final states to be obtained, in which different words coexist forever in the system \cite{baronchelli_ng_trans}. In the second scenario, on the other hand, testing the details of the post-success updating determine it is possible to show that the same global phenomenology produced by the NG and the minimal NG can be obtained with even simpler rules \cite{baronchelli2011role}. In the coming section we analyze the main points of these variants to show the importance and the potential of a detailed scrutiny of all aspects of the rules defining a model, even though when might appear to be already very simple.

\subsubsection*{A consensus-polarization transition in the minimal NG}

In the minimal NG, after a successful interaction the agents delete all the words except the one they have just agreed upon. In \cite{baronchelli_ng_trans} a parameter was introduced, $\beta$, stating the {\it probability} of this update. All the remaining rules remain unchanged, and the usual model is recovered for $\beta = 1$.  Thus, $\beta$ mimics an  irresolute attitude of the agents to
make decisions. Interestingly, the new negotiation process displays a non-equilibrium phase transition from an
absorbing state, in which all agents reach a consensus, to an active stationary state characterized either by polarization or fragmentation in clusters of agents with different opinions. Single agents keep negotiating, and they update their inventories accordingly, but the statistical abundances of the surviving words are stable.

Without entering into the details, it is worth stressing that it is possible to identify the critical value $\beta_c$ analytically, for which if $\beta>\beta_c$ the final state is always consensus, while if $\beta<\beta_c$ two different words will survive \cite{baronchelli_ng_trans}. Moreover, it can be proven, and has been observed, that the transition occurs also when the population is embedded in complex networks, which is remarkable since in many other cases disordered topologies wipe out similar transitions \cite{rmp2009}. Finally, it is interesting to note that the consensus-to-two-words transition is just the first of a series of similar discontinuities. Figure \ref{fig_transitions} shows indeed that lowering $\beta$ it is possible to tune the number of words that will survive asymptotically in the system.

\subsubsection*{The role of feedback}
This section also takes the success rule  as its focus, but a different aspect is analyzed. The fact that both agents undergo the very same operation (i.e. shrink their inventories to the same unique word) underlies the existence of a feedback between the two. In the original formulation the feedback occurs through an outside world, with the hearer pointing to the object he would associate with the received word. The speaker would then point on its turn to the right object, and both individuals would immediately know whether the game was a success or a failure \cite{Steels1996}. In the minimal NG, however, the feedback simply consists in the hearer informing the speaker that he or she too has the transmitted word. In case of failure, on the other hand, no feedback is needed. 

In \cite{baronchelli2011role} we have investigated what happens when only one of the agents updates his inventory after a successful interaction. The result is that the situation changes dramatically depending on whether the update is performed by the hearer or the speaker only, which are cases referred to as Hearer Only NG (ho-NG) and Speaker-Only NG (so-NG) respectively. In particular,  the ho-NG yields a scaling of the convergence time with the population size that is the same as the one observed in the usual NG. The so-NG, on the other hand, is significantly slower. The reason beyond this difference can be understood analytically in the light of the generalized $\beta$-model discussed above, showing that it spontaneously falls in the critical regime of the generalized model, i.e. that, for the so-NG, $\beta_c=1$.

The result concerning the ho-NG is interesting, too. Indeed, the fact that the ho-NG behaves substantially in the same way as the usual NG implies that the hearer's feedback to the speaker is not crucial, and opens the way to the implementation of straightforward broadcasting protocols on networks, in which a speaker can speak at the same time to all of his neighbors without having to bother about receiving any feedback. Crucially this strategy allows for a faster convergence of the dynamics \cite{baronchelli2011role}. It must be noted, however, that the fact that the ho-NG and the NG behave in the same way as far as the scaling with the system size of the relevant quantities is concerned holds in the framework of the minimal NG only. It is indeed likely to be a consequence of the fact that homonymy is not taken into account. In fact, feedback remains a fundamental ingredient of any language game, as devised by Wittgenstein \cite{wittgenstein53english}.

\section{Conclusion}

The Naming Game is a fundamental model in semiotic dynamics, addressing possibly the most basic issue in the tremendously tough problem of language evolution. The minimal Naming Game that we have partially reviewed in this chapter is the result of a further simplification effort put forth to favor more systematic and deep investigations of its dynamics. Thanks to particularly transparent rules, that yet are able to reproduce qualitatively the same overall dynamics observed in the NG, the minimal version can indeed be studied in great detail resorting in the conceptual and technical tools developed in statistical physics and complex systems science. Moreover, remarkably, the minimal model has been shown to correctly describe experimental results on the spontaneous emergence of social conventions \cite{centola2015spontaneous}.

We have discussed the population-scale dynamics of the minimal NG in unstructured as well as structured populations, pointing out the role played by the underlying topology. We have seen that finite connectivity implies a finite memory requirement for the agents, while the small-world property yields a faster convergence. Finally we have looked at two slightly modified versions of the model bearing a consensus-to-polarization transition and some interesting insights on the role of feedback, respectively.

The examples we have analyzed represent only a subset of the studies triggered by the definition of the minimal NG, but hopefully they give an idea of the potentiality of the fruitful exchange between the fields of semiotic dynamics and the statistical physics approach to complex systems. This method has also been profitably applied to more complex issues, such as categorization. In this context, the Category Game model \cite{cg_pnas}, which is literally built on top of the minimal NG, has proven to be able to reproduce experimental data concerning color naming systems \cite{baronchelli10,loreto2012origin,baronchelli2015individual}. Different research avenues remain open for the future, ranging from addressing more complex problems such as the emergence compositionality \cite{tria2012naming,roberts2015communication} to understanding the nature of language change \cite{cuskley2014internal,colaiori2015general}, and there is consensus among the researches in different disciplines on the substantial contribution that the complex systems approach will continue to provide \cite{frankfurt}.

\subsubsection*{Acknowledgments} The author is indebted to V. Loreto, A. Barrat, L. Dall'Asta, A. Puglisi and L. Steels for their major contribution to the study of the minimal Naming Game and for the subsequent research projects.  

\vspace{2cm}


\end{document}